\documentclass[11pt]{article}
\usepackage{bm,amsfonts,amssymb,amsthm,amsmath,gensymb}
\usepackage{mathrsfs}
\usepackage{graphicx,color}
\usepackage{CJK}
\usepackage{indentfirst,enumerate}

\newcommand{\md}{\mathrm{d}}

\definecolor{orange}{rgb}{1,0.5,0}
\definecolor{rb}{rgb}{1,0,1}
\allowdisplaybreaks[1]

\begin{document}
\title{Computing optimal interfacial structure of modulated phases}
\author{Jie Xu$^1$, Chu Wang$^{2}$, An-Chang Shi$^3$ and Pingwen Zhang$^1$\footnote{Corresponding author}\vspace{12pt}\\
\small
$^1$LMAM \& School of Mathematical Sciences, Peking University, Beijing 100871, China\\
\small
$^2$Program in Applied and Computational Mathematics, Princeton University, \\
\small Princeton, New Jersey 08544, USA\\
\small
$^3$Department of Physics and Astronomy, McMaster University, \\
\small Hamilton, Ontario L8S4M1, Canada\\
\small
Email: rxj\_2004@126.com, chuw@math.princeton.edu, shi@mcmaster.ca, pzhang@pku.edu.cn}
\date{\today}

\maketitle

\begin{abstract}
We propose a general framework of computing interfacial structures between 
two modulated phases. Specifically we propose to use a computational box consisting of two half spaces, 
each occupied by a modulated phase with given position and orientation. 
The boundary conditions and basis functions are chosen to be commensurate with  
the bulk structures. It is observed that 
the ordered nature of modulated structures stabilizes the interface, 
which enables us to obtain optimal interfacial structures by searching 
local minima of the free energy landscape. 
The framework is applied to the Landau-Brazovskii model to investigate interfaces between modulated phases with different relative positions and orientations. Several types of novel complex interfacial structures are obtained from the calculations. 

{\bf Keywords:} Interface; Modulated phase; Metastable state; Compatibility; Landau-Brazovskii model. 
\end{abstract}

\section{Introduction\label{Intro}}
Interfaces are transition regions connecting two different materials, 
two different phases of the same material, or two grains of the same phase 
with different orientations (grain boundaries). Interfacial regions are where 
the symmetries and patterns of the ordered structures are interrupted. 
Therefore, the structure of interfaces greatly affects the mechanical, 
thermal and electrical properties of a material. Interfaces are frequently 
encountered as planar defects in phase transitions. In particular, 
the strength and conductivity of a material depend strongly on the 
distribution and morphology of grain boundaries. 
In first-order phase transitions, the interfacial properties 
play an important role in the nucleation-growth process. 

Theoretical discussions of interfaces usually start from the coexistence of two 
homogeneous phases, namely the order parameters are spatially uniform 
in equilibrium. If the contribution of inhomogeneity is included in the 
free energy, the interfacial structure becomes an intrinsic property of the 
energy functional. 
A simple but widely-used energy functional of this model system is proposed by 
Cahn and Hilliard \cite{Cahn_Hilliard_1}. 
Since interfaces are non-equilibrium structures with long relaxation time, 
two different points of view can be held. 
One regards the interface as a metastable state and its morphology is 
considered as a local minimizer of the free energy under certain constraints. 
The minimization approach is able to reach full relaxation and resolves 
the interfacial structure. This approach has been applied to two-component 
fluid interfaces to study the thickness and shape in various circumanstances 
\cite{Cahn_Hilliard_1,Cahn_Hilliard_3,Oxtoby1988,JChP1996,JChP2001}, 
as well as isotropic-nematic interfaces in liquid crystals 
\cite{Oxtoby1988,I_N_1,I_N_2}. 
An alternative approach is to treat the interface as a transient state and 
focuses on its dynamics, based on the free energy, in some complex processes 
(see \cite{DynCH_1,DynCH_2} for two examples built on the Cahn-Hilliard energy).
The dynamical approach enables the study of the dynamical evolution of 
interfaces. 

Interfaces between modulated phases have unique features. Because of spatial 
modulation, the interfacial profile depends on the relative position and 
orientation of the phases. Also interfaces may exist between two grains 
of the same phase, i.e. grain boundaries. These features make
it extremely interesting to study the mechanism of how two modulated 
structures are connected, which is very helpful to understanding the origin of 
epitaxial relationship and the anistropic nucleation. 
Therefore, it is important to investigate the morphology of a single interface 
using the minimization approach. In previous studies, the minimization approach 
has been used successfully in the tilted grain boundaries of 
the lamellar phase \cite{PRL1997,JChemP1997,PRE2000,JChP2002} 
and the bcc phase \cite{PRE2009}, 
and twist grain boundaries of several cubic phases \cite{JChemP2009}. 
Some works use dynamical apporach \cite{PRE2004,Macro2005,PhTrRSocA2011}, 
but it usually generates several interfaces because there is limitation in 
choosing boundary conditions, which we will explain later in detail. 
From the view of computation, dynamical approach is more time-consuming, 
while in the minimization approach fast optimization algorithms can be used. 
In what follows, our discussion is limited to the minimization approach. 

To convert a non-equilibrium interface into a metastable state, we need some 
anchoring conditions. 
Let us explain the anchoring conditions using a planar liquid-vapor interface 
as an example, 
where the density $\phi$ can be viewed as varying only in the $x$-direction. 
Suppose that the density of the liquid is $\phi_1$, 
and that of the vapor is $\phi_2$. 
The density far away from the interface shall be identical to the bulk values, 
\begin{equation}\label{bnd_anch}
\phi(-\infty)=\phi_1,\ \phi(+\infty)=\phi_2. 
\end{equation}
Note that these conditions do not determine the location of the interface. 
If we want to fix it, say, at $x=0$, an extra constraint is needed. 
A possible constraint is to choose an interval $[-L,L]$ and 
fix the total number of molecules in it, 
\begin{equation}\label{num_cons}
\int_{-L}^L\md x~\phi(x)=N. 
\end{equation}
The constraint can easily be extended to the study of interfaces of other 
shapes on substrate or with external forces \cite{JChP1996,JChP2001}. 
The conditions (\ref{bnd_anch}) and (\ref{num_cons}) are both physical, 
and are sufficient to pose the interfacial profile as a minimization problem. 

For interfaces between two modulated phases, the anchoring conditions are not 
straightforward to specify. We need to consider the relation between the 
conditions and the bulk profiles. 
Since the anchoring conditions actually specify the function space in which 
the minimization problem is solved, we may evaluate how 
good the conditions are by comparing the function space with the bulk profiles. 
If the bulk profiles are included in the function space, we say that the 
anchoring conditions are compatible. The condition (\ref{bnd_anch}) is 
compatible, and it becomes incompatible if we set $\phi(+\infty)$ to a value 
other than $\phi_2$. 
We believe that compatibility would be more significant for modulated phases, 
because we need to anchor the phases with given relative position and orientation. 
Besides the compatibility, the anchoring shall include as few artificial 
restrictions as possible to enable the free relaxation of the interface. 

In previous works, however, the compatibility and the reduction of artificial 
effects are usually not considered simultaneously. 
As a simple extension of the disordered interfaces, 
some works construct the interface profile by a mixing ansatz, 
which is the direct weighted combination of two bulk phase profiles: 
\begin{equation}\label{connection}
  \phi(x,y,z)=(1-\alpha(x))\phi_1(x,y,z)+\alpha(x)\phi_2(x,y,z)
\end{equation}
where $\phi_k$ are profiles of two bulk phases, and $\alpha(x)$ is a 
smooth monotone function satisfying $\alpha(-\infty)=0,\ \alpha(+\infty)=1$. 
This method proves to be convenient and effective as shown in the literatures 
\cite{Oxtoby1988,JPSJ2007,SoftM2011}. 
But this artificial approach may exclude the possibility of complex interfacial 
structures as we will present later in this paper. 
Another convenient method is to fill a large cell with modulated structures and 
let the interface relax itself (see, for example, \cite{PRE2004}). 
However, besides the computational challenges, the boundary conditions commonly 
used and the bulk structures are incompatible. 
Although large-cell computations have brought some beautiful results in 
\cite{PRL1997}, where considerable efforts are made to fit the bulk structures 
with the boundary, the incompatibility may easily generate many interfaces 
together, like Fig. 5.20 in \cite{Fredrickson}. 
On the other hand, it may alter the position and orientation of 
the bulk structures from the desired values, 
or even destroy the bulk structures (see the examples in \cite{Macro2005}). 

There have been attempts to balance between anchoring bulk structures and 
reducing the artificial effects. 
In the study of kink \cite{JChemP1997} and T-junctions \cite{JChP2002} in 
lamellar grain boundaries, a set of basis functions are carefully chosen to 
retain symmetry properties. However, the basis functions are only partially 
compatible with the lamellar structure, 
and they cannot be easily extended to other structures. 
Therefore, we aim to propose compatible anchoring conditions with universal 
applicability and no artificial effects. 

When investigating an interface between different phases, a mechanism is 
required to prevent it from moving gradually towards the phase with higher 
energy density. 
We may choose parameters to equalize the energy densities, 
but it is difficult to realize in computation. 
Intuitively, we need to propose an analog of the constraint (\ref{num_cons}). 
However, we observe in the computation that constraints of such kind 
are not necessary if the bulk phases are modulated. 
The interface will be pinned at a locally optimized position as long as 
the energy densities of the two phases do not differ too much. 
Thus we may let the interface freely relax itself during the computation 
instead of intervening in the process. 

In this work, we propose a general framework for the computation of 
interfacial structures between modulated phases. 
We consider two phases or grains, large in size, connected via an interfacial region. 
We choose basis functions compatible with the bulk structures in two directions 
parallel to the contact plane, and use a compatible boundary anchoring 
analogous to (\ref{bnd_anch}) in the direction vertical to the plane. 
The computational setting is applicable to any modulated phases, 
and is well-posed as a minimization problem for phases with energy difference. 
By this setting, we can take the advantage of full relaxation of the system 
and fast optimization methods. 
We will apply this framework to the Landau-Brazovskii model to illustrate 
the above features. 
Several lamellar-gyroid and cylindral-gyroid interfaces with 
different relative positions and orientations are examined, 
and some complex structures are acquired. 
The rest of the paper is organized as follows. 
In Sec. \ref{alg}, the computational framework is described, 
and its well-posedness is illustrated. Some interfacial structures 
in the Landau-Brazovskii model are presented in Sec. \ref{result}. 
Finally we summarize the paper in Sec. \ref{concl}. 
Some details of numerical method are given in Appendix. 

\section{Computational framework\label{alg}}
\subsection{Compatible anchoring conditions\label{extn}}
When two ordered structures contact each other, 
their spatial and orientational relations are essential variables. 
Let us place the two phases in two half-spaces separated by the plane $x=0$. 
Denote the two phases by $\alpha$ and $\beta$ respectively. 
Both phases can be rotated and displaced, represented by the orthogonal rotation
matrices $\mathcal{R}_{\alpha},\ \mathcal{R}_{\beta}$, 
and the displacement vectors ${\bm d}_{\alpha},\ {\bm d}_{\beta}$. 
Each of these four quantities has three degrees of freedom. 
Note that the relation of two phases remains unchanged if they are rotated 
together round the $x$-axis or shifted together in the $y$-$z$ plane, 
so there are nine independent degrees of freedom in total. 

Assume that the density profile, or order parameter, of a phase can be 
expressed by a scalar function $\phi$. 
More specifically, we assume that $\phi$ is periodic, 
which can be written as 
\begin{equation}
  \phi(\bm{r})=\sum_{\bm{k}\in\mathbb{Z}^3}\phi_{\bm{k}}
  \exp\left(i\sum_{j=1}^3 k_j\bm{b}_j\cdot {\bm r}\right). \label{prof_per}
\end{equation}
If the phase is rotated by $\mathcal{R}$, then shifted by $\bm{d}$, the profile 
becomes
\begin{equation}
  \phi(\bm{r};\mathcal{R},\bm{d})=
  \phi(\mathcal{R}^T(\bm{r}-\bm{d}))=\sum_{\bm{k}\in\mathbb{Z}^3}
  \phi_{\bm{k}}\lambda_{\bm{k}}
  \exp\left(i\sum_{j=1}^3 k_j(\mathcal{R}\bm{b}_j)\cdot {\bm r}\right), 
\end{equation}
where $\lambda_{\bm{k}}=\exp(-i\sum_{j=1}^3 k_j\bm{d}^T\mathcal{R}\bm{b}_j)$. 

As mentioned above, we aim to propose anchoring conditions with compatibility. 
First we check the $y$- and $z$-directions. 
Consider the projection of profile $\phi_{\alpha}$ and $\phi_{\beta}$ onto the plane $x=x_0$. 
Denote ${\bm r}'=(y,z)$. Then we have 
$$
\phi_{\alpha}(x_0,\bm{r}';\mathcal{R}_{\alpha},\bm{d}_{\alpha})=
\sum_{\bm{k}\in \mathbb{Z}^3}\phi_{\alpha\bm{k}}\tilde{\lambda}_{\alpha\bm{k}}\exp\left(i 
\sum_{j=1}^{3}k_j\bm{b}_{\alpha j}'(\mathcal{R}_{\alpha})\cdot \bm{r}'\right), 
$$
where ${\bm b}_{\alpha j}'(\mathcal{R}_{\alpha})$ denotes the $y$ and $z$ 
components of $\mathcal{R}_{\alpha}{\bm b}_{\alpha j}$, and
$\tilde{\lambda}_{\alpha\bm{k}}=\lambda_{\alpha\bm{k}}\exp\big(ix_0\sum_{j=1}^3
k_j(\mathcal{R}_{\alpha}\bm{b}_{\alpha j}')_1\big)$. 
It becomes a quasiperiodic function in the plane. 
And for $\phi_{\beta}$, we have 
$$
\phi_{\beta}(x_0,\bm{r}';\mathcal{R}_{\beta},\bm{d}_{\beta})=
\sum_{\bm{k}\in \mathbb{Z}^3}\phi_{\beta\bm{k}}\tilde{\lambda}_{\beta\bm{k}}\exp\left(i 
\sum_{j=1}^{3}k_j\bm{b}_{\beta j}'(\mathcal{R}_{\beta})\cdot \bm{r}'\right). 
$$
Then a natural choice will be the set of quasiperiodic function
\begin{equation}
\mathcal{F}=\left\{f({\bm r}'):f({\bm r}')=\sum_{{\bm k},\bm{l}\in\mathbb{Z}^{3}}f_{{\bm k}}
\exp\left(i\sum_{j=1}^{3} \big(k_j{\bm b}_{\alpha j}'(\mathcal{R}_{\alpha})+l_j{\bm b}_{\beta j}'(\mathcal{R}_{\beta})\big)\cdot {\bm r}'\right) \right\}. 
\label{basis}
\end{equation}
In general, a quasiperiodic function is a projection of a higher-dimensional 
periodic function onto a lower-dimensional linear subspace. 
Actually, this choice is also valid if $\phi$ itself is quasiperiodic, 
namely to substitute the sum over $1\le j\le 3$ with $1\le j\le m$. 
In special cases where $k_j{\bm b}_{\alpha j}'
(\mathcal{R}_{\alpha})+l_j{\bm b}_{\beta j}'(\mathcal{R}_{\beta})$ 
lie on a 2D lattice, $\mathcal{F}$ is reduced to a set of periodic functions, 
indicating that two phases, with their relative position and orientation 
determined, have common period in the $y$-$z$ plane. 
The current work will focus on these special cases. 
Despite we do not consider the general quasiperiodic cases currently, 
interfaces of this type have been observed (see Fig. 24 of \cite{HREM_GB}). 

\begin{figure}
  \centering
  \includegraphics[width=0.7\textwidth,keepaspectratio]{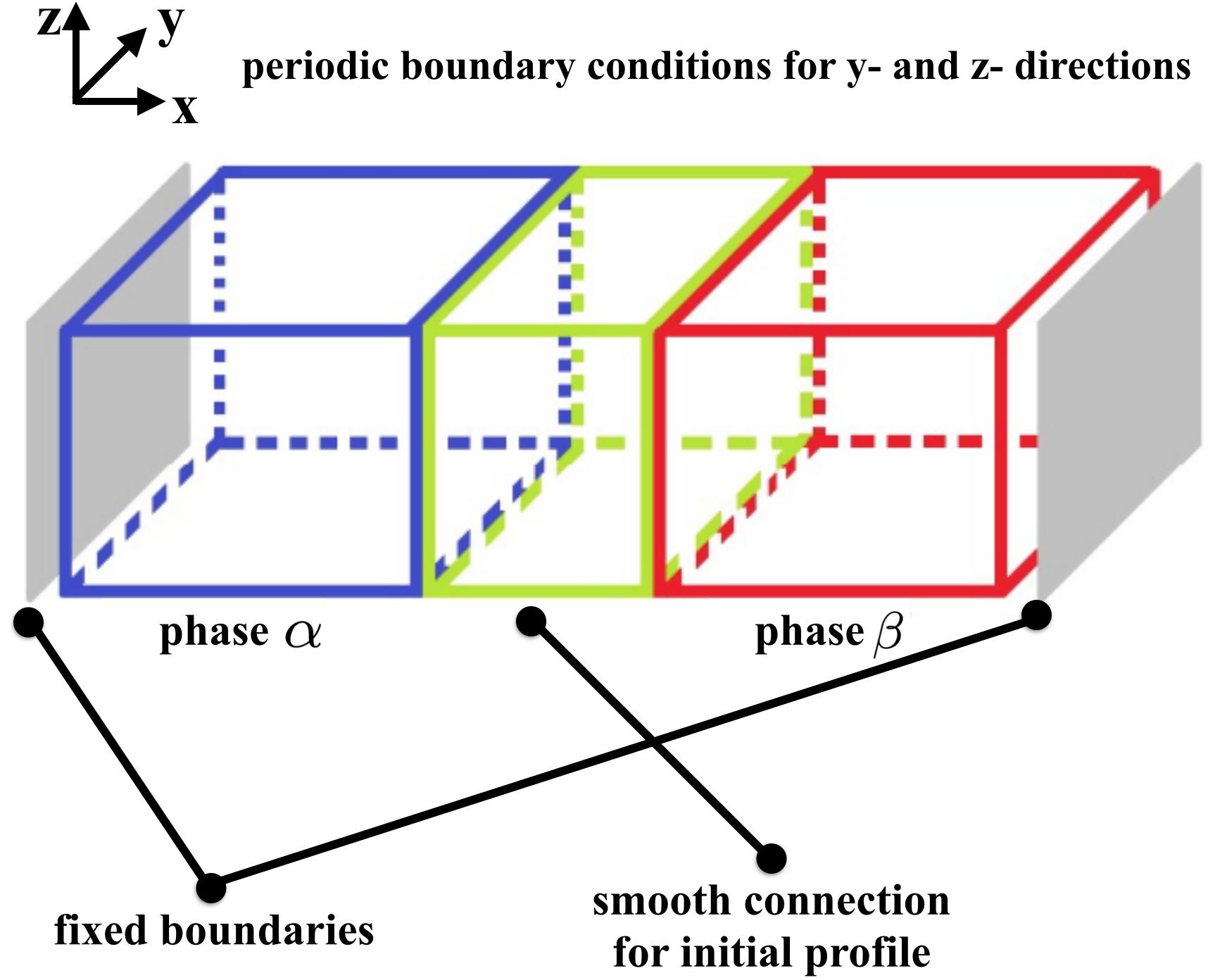}
  \caption{Schematic of the setting of interface problem. }\label{setting}
\end{figure}
In the $x$-direction, we select a length $L$ and set $\phi$ outside 
$[-L,L]$ equal to the bulk value. 
Here $L$ is large enough to contain the transition region. 
Such setting will induce some anchoring conditions at $x=\pm L$ 
dependent on the energy functional. If a Laudau-type energy functional is 
used, these conditions can usually be determined by smoothness requirements 
of the density profile $\phi$. For example, if $\phi$ is $C^k$, 
then $\phi,\ldots,\nabla^k \phi$ shall be fixed to bulk values at $x=\pm L$. \
It should be noted that in some other models 
boundary conditions are not directly imposed on the density profile. 
For example, in self-consistent field theory (we refer to \cite{Fredrickson} 
for details), the profile is calculated through a propagator $q$, 
on which boundary conditions are imposed. 
In this case the anchoring conditions can be used on $q$. 

To initialize the density profile for computation in our frame work, 
we use the setting in Fig. \ref{setting}.
We first choose a common period in the $y$- and the $z$- direction for 
phase $\alpha$ and $\beta$, 
and then fill in the bulk profiles and anchor both ends of the region.
To obtain a smooth initial value, 
the density profile in the middle region is set as the 
convex combination of the bulk densities as in (\ref{connection}). 

\subsection{Existence of local minima}
\begin{figure}
  \centering
  \includegraphics[width=0.5\textwidth,keepaspectratio]{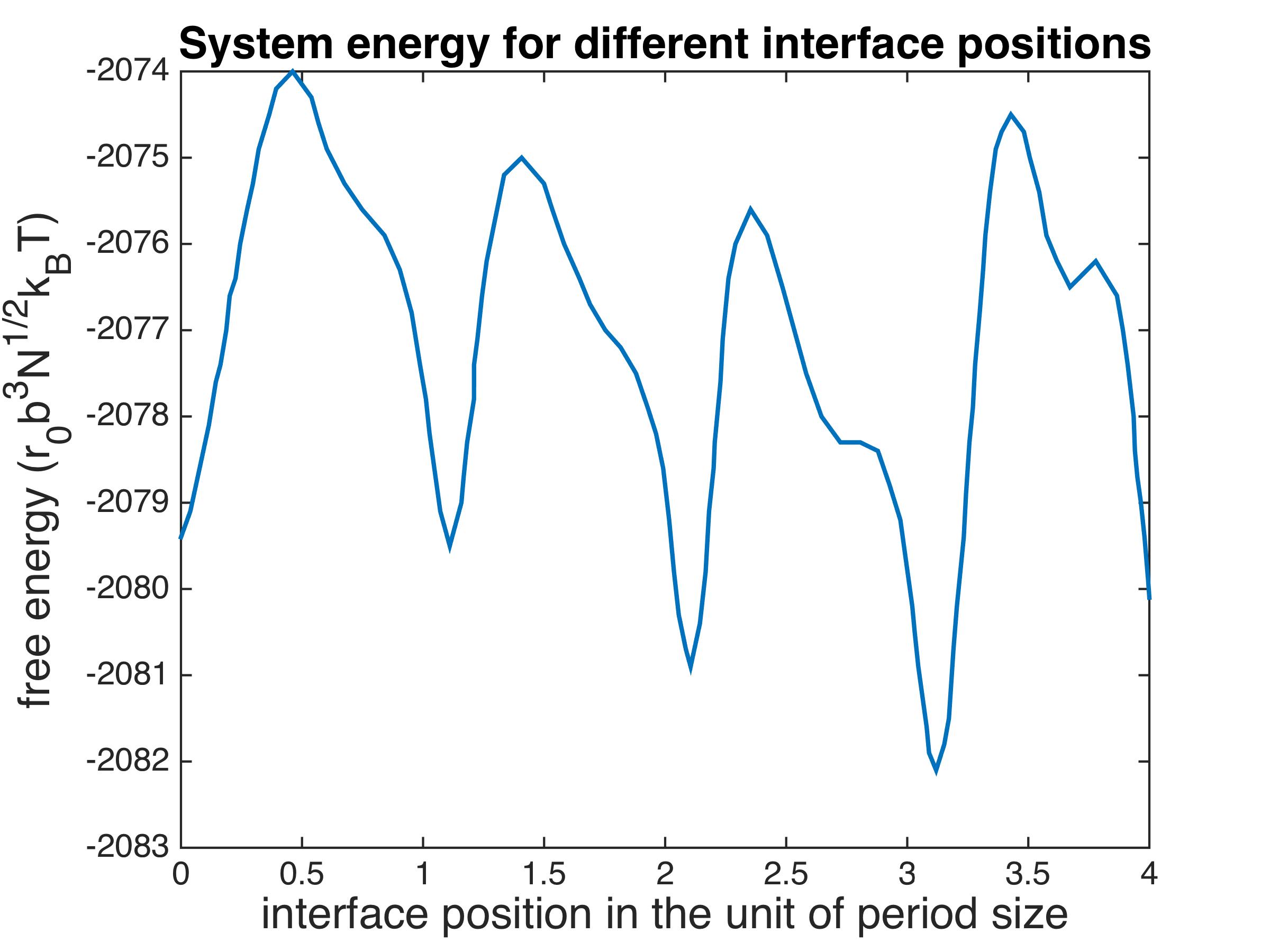}
  \caption{Energy landscape in the movement of cylindral-gyroid interface in 
    the Landau-Brazovskii model (\ref{LB}) within a single period. 
    The parameters in (\ref{LB}) are chosen as $\xi^2=1.0,\ \tau=-0.383,\ 
    \gamma=0.3$. 
    The location of interface is calculated from the distance between the 
    points on the minimal energy path \cite{str0,simpstr}. 
    The distance is measured by the unit cell of gyroid, 
    which is rescaled to four (the upper bound of the $x$-axis). 
    For the morphology at local minima, see Fig. \ref{match}. 
    Although we choose different parameters, the morphology is identical. 
}
  \label{mep}
\end{figure}
Before solving the optimization problem, 
the well-posedness of this setting should be discussed.
Specifically, we need to demonstrate the existence of local minima 
in our setting. 
It is obvious that for disorder-disorder interface, 
if there exists a difference in the bulk energy densities, 
no matter how small it is, 
the interface would move continuously to the one with higher energy density. 
In this case, the total energy can be written as 
\begin{equation}\label{eng0}
E=f_{\alpha}V_{\alpha}+f_{\beta}V_{\beta}+\gamma S 
\end{equation}
where $f$ is the free energy density, $V$ is the volume of each phase, 
$S$ is the area of interface, and $\gamma$ is the interfacial energy density. 
The isotropy of two phases along the $x$-direction makes $\gamma$ independent 
of interface location. Suppose $f_{\alpha}<f_{\beta}$, 
then $E$ is monotone decreasing when $V_{\alpha}$ increases, 
driving the interface to the phase $\beta$. 
For homogeneous phases, a constraint like (\ref{num_cons}) is usually easy to 
propose to fix the volume fraction of each phase. 
Nevertheless, it is difficult to extend to modulated phases. 

Fortunately, we can take advantage of the anisotropy in modulated phases. 
The anisotropy implies that $\gamma$ is no longer constant. 
Intuitively we write $\gamma=\gamma(x)$ as a function of interface location, 
and still write the total energy as (\ref{eng0}). 
If $\gamma(x)$ varies more drastically than the bulk energy difference, 
some local minima exist, towards which we may let the interface relax. 
Such condition is easier to be attained when 
the energy difference between two phases is not large. This can be achieved 
when model parameters lie in a region near the binodal line. 
In fact, if the parameters are far away from the binodal line, 
usually it is difficult for the interfaces to exist for long time, 
for the phase with higher energy density is more probable to decompose 
because of fluctuations. 

A typical energy landscape is like what is plotted in Fig. \ref{mep}, where 
we compute the cylindral-gyroid interfaces in the Landau-Brazovskii model 
(\ref{LB}). It shows clearly that the energy decreases in a wavy manner 
as the interface moves. 
Four local minima are found within a single period along the $x$-direction, 
and they are connected with the minimum energy path computed using string 
method \cite{str0,simpstr}. 
The morphology of interfaces at local minima is drawn in 
Fig. \ref{match} (where we draw the interfaces within two periods). 

We note that only a few works study interfaces between different 
modulated phases with such a weak anchoring condition. 
The reason could be the lack of knowledge of the existence of 
local minima produced by modulated structures. 
Although in the current work it is only confirmed in one model, 
we believe that this phenomenon is generic, 
and that the clarification of the energy landscape would be helpful to 
the computation of interfacial structure in other systems. 

\section{Application to the Landau-Brazovskii model\label{result}}
In this section, we apply our framework for interface to the 
Landau-Brazovskii (LB) model. We first present 
the cylindral-gyroid and lamellar-gyroid interfaces 
in epitaxially matching cases, in which the local minima are shown clearly. 
Furthermore, a few novel examples of non-matching cases are given 
to show the effectiveness of the framework. 

\subsection{The free energy}
\begin{figure}
  \centering
  \includegraphics[width=0.5\textwidth,keepaspectratio]{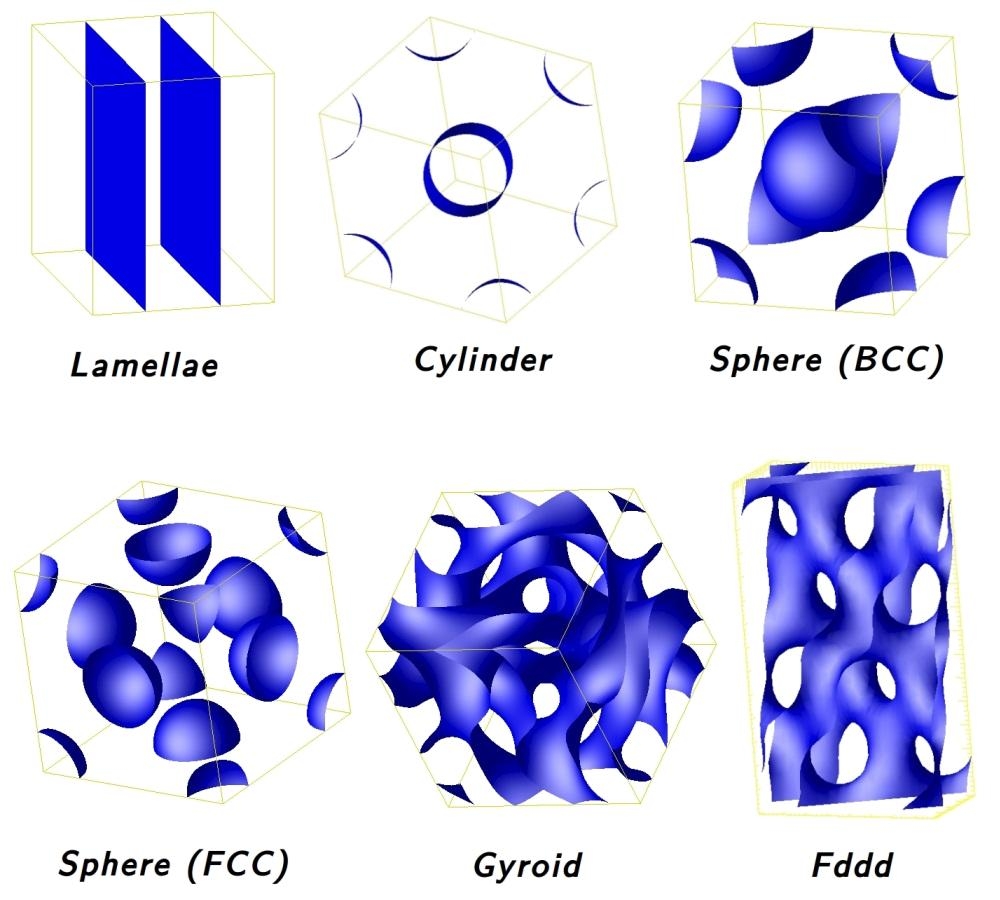}
  \caption{Bulk phases of the Landau-Brazovskii model.}\label{phases}
\end{figure}
The LB model is a free energy functional appropriate for weak crystallization 
\cite{LB,JChP87,PhRep1993}. This model can be viewed as a generic 
model for modulated phases occurring in a variety of physical and chemical 
systems, and its form is similar to many Landau-type free energy functionals 
for different kinds of materials. In addition, the LB model can describe frequently 
observed patterns, including lamellar(L), cylindral(C), spherical and gyroid(G) 
structures. 
Therefore our results could reveal properties of interfaces in a wide range of 
systems. 
In its scaled form, the LB free energy density is given by 
\begin{equation}\label{LB}
F[\phi]=\frac{1}{V}\int_{\Omega}\md{\bm r}\{\frac{\xi^2}{2}[(\nabla^2+q_0^2)\phi]^2\}
+\frac{\tau}{2}\phi^2-\frac{\gamma}{3!}\phi^3+\frac{1}{4!}\phi^4, 
\end{equation}
where $q_0=1$ is the critical wavelength, $\xi,\ \tau,\ \gamma$ are 
phenomenological parameters, and $\phi$ is conserved, 
$$
\int\md\bm{r}\phi=0. 
$$ 
The parameters can be determined by measurable parameters in some cases. 
An example is the system of A-B diblock copolymer, 
in which these parameters are derived from $\chi N$ and $f$, 
where $\chi N$ is a normalized parameter characterizing the segregation of 
two blocks, and $f$ is the fraction of block A. 
The phases in the LB model can be easily recognized by the isosurface of $\phi$, 
drawn in Fig. \ref{phases}. For the interfaces, we will also draw the 
isosurface to reveal their structures. 

\subsection{Boundary conditions}
Bulk values of $\phi$ are needed for setting initial and boundary values of the problem. 
They are obtained by minimizing (\ref{LB}) with periodic boundary condition in 
all three directions. The existence of second-order derivatives 
in the energy functional requires $\phi$ and $\nabla \phi$ to be fixed at 
$x=\pm L$. These values can be easily computed with bulk profiles. 

When computing bulk profiles, the period lengths should be optimized as well.
This is because the size of the cell also affects the energy density of 
the system. Refined computation shows that for different phases, there is 
slight distinction between the optimal size of unit cell \cite{LBPhase}. 
When two phases coexist, however, it is usually observed experimentally that 
two unit cells match each other \cite{PRL1994, Ma94, JPCB2003, JACS2009}. 
To capture the interfacial structure in our framework, 
we slightly stretch the two bulks to let them have common period lengths. 
Then our framework for the interface could be applied directly. 
Such approximation leads to a small amount of the bulk energy increase, 
but we will adopt it as it is consistent with experimental results. 
It should be pointed out that besides the period matching, different 
modulated structures have certain preferences in orientation when they coexist, 
namely the rotations $\mathcal{R}_{\alpha},\ \mathcal{R}_{\beta}$ and the shifts 
${\bm b}_{\alpha},\ {\bm b}_{\beta}$ prefer certain values. 
Such epitaxial relationships are also noted in the experimental works 
mentioned above, and are studied in \cite{SoftM2011} extensively. 
We will examine such epitaxies as well as less optimal matching cases 
for the interfacial systems. 

The profiles of L, C and G are calculated with the common period 
$2\sqrt{6}\pi\times 2\sqrt{6}\pi\times 2\sqrt{6}\pi$, which is almost accurate 
for L and C, while about $4\%$ smaller for G. 
The number of meshes used in a unit cell is $32\times 32\times 32$. 
Details of discretization and optimization method are given in Appendix, 
in which acceleration techniques are included. 

\subsection{C-G and L-G interfaces}
\begin{figure}
  \centering
  \includegraphics[width=0.8\textwidth,keepaspectratio]{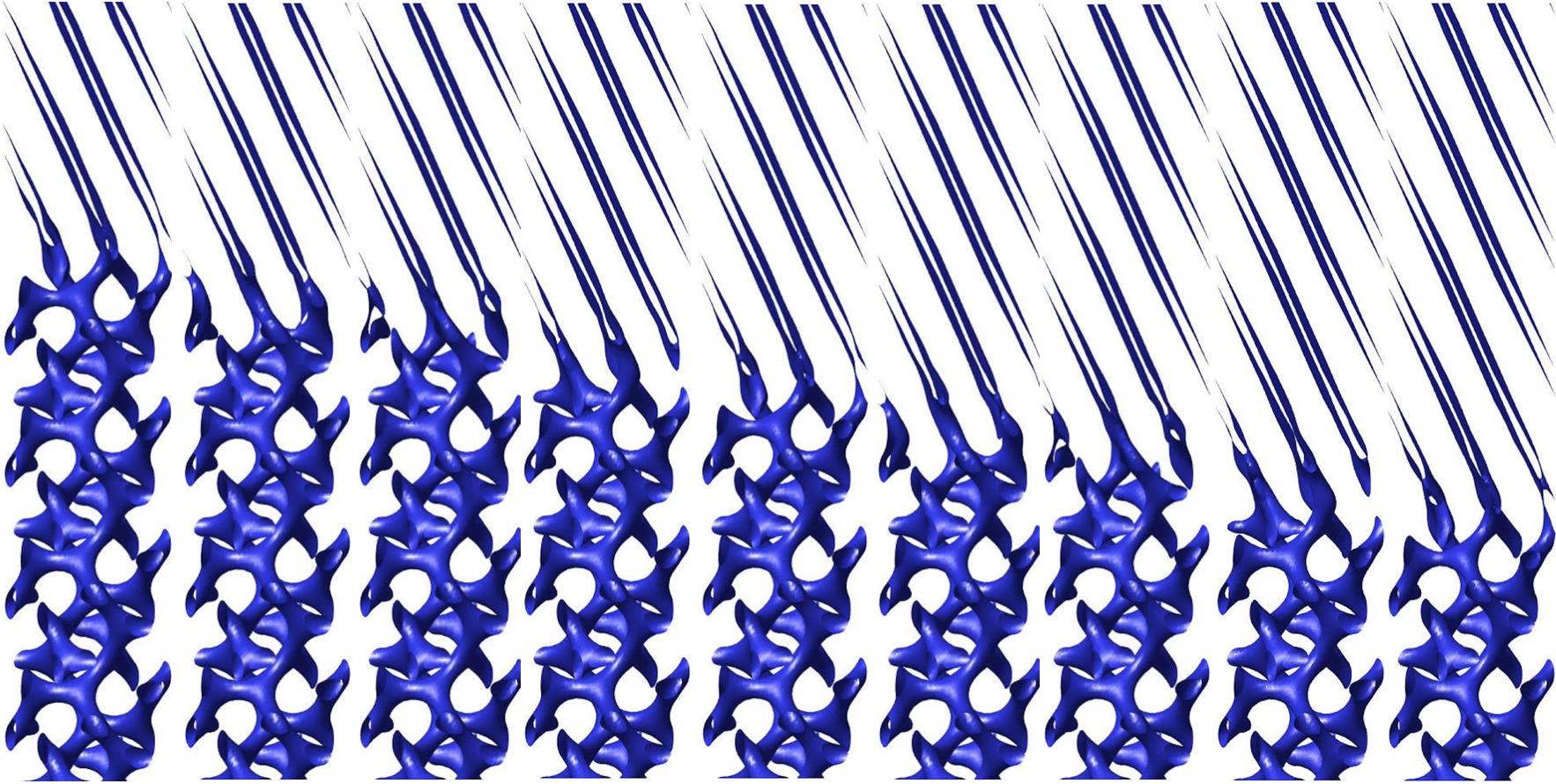}\\\vspace{10pt}
  \includegraphics[width=0.8\textwidth,keepaspectratio]{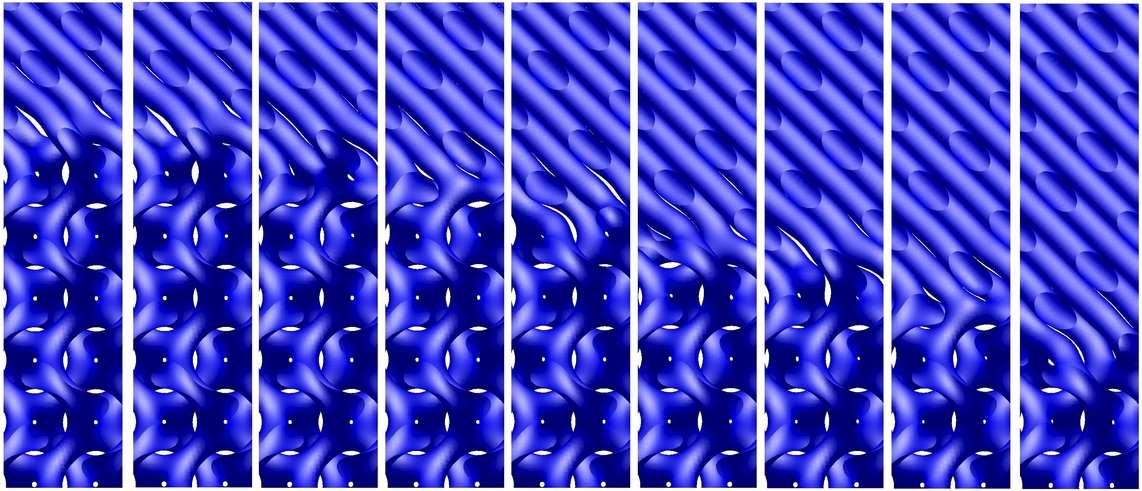}
  \caption{L-G and C-G interfaces at local minima (within two periods): 
    the epitaxially matching case. }
  \label{match}
\end{figure}
\label{numres}
We start from the cases in which two phases are epitaxially matched: 
in the lattice of G, the layer of L parallels to the plane $(11\bar{2})$ and 
the hexagonal lattice of C lies in the plane $(111)$. 
The results in Fig. \ref{match} are calculated with 
$\xi^2=0.0389,\ \gamma=0.0681$; $\tau=-0.0121$ for C-G interfaces 
and $\tau=-0.0159$ for L-G interfaces. 
Fig. \ref{match} presents C-G and L-G interfaces at local energy minima. 
Both C-G and L-G interfaces show four local minima within one period, 
and when moving a full period, the interfacial structures reappear. 
We are also able to capture how the interface moves from 
the ones at discrete locations. 

\begin{figure}
  \centering
  \includegraphics[width=0.48\textwidth,keepaspectratio]{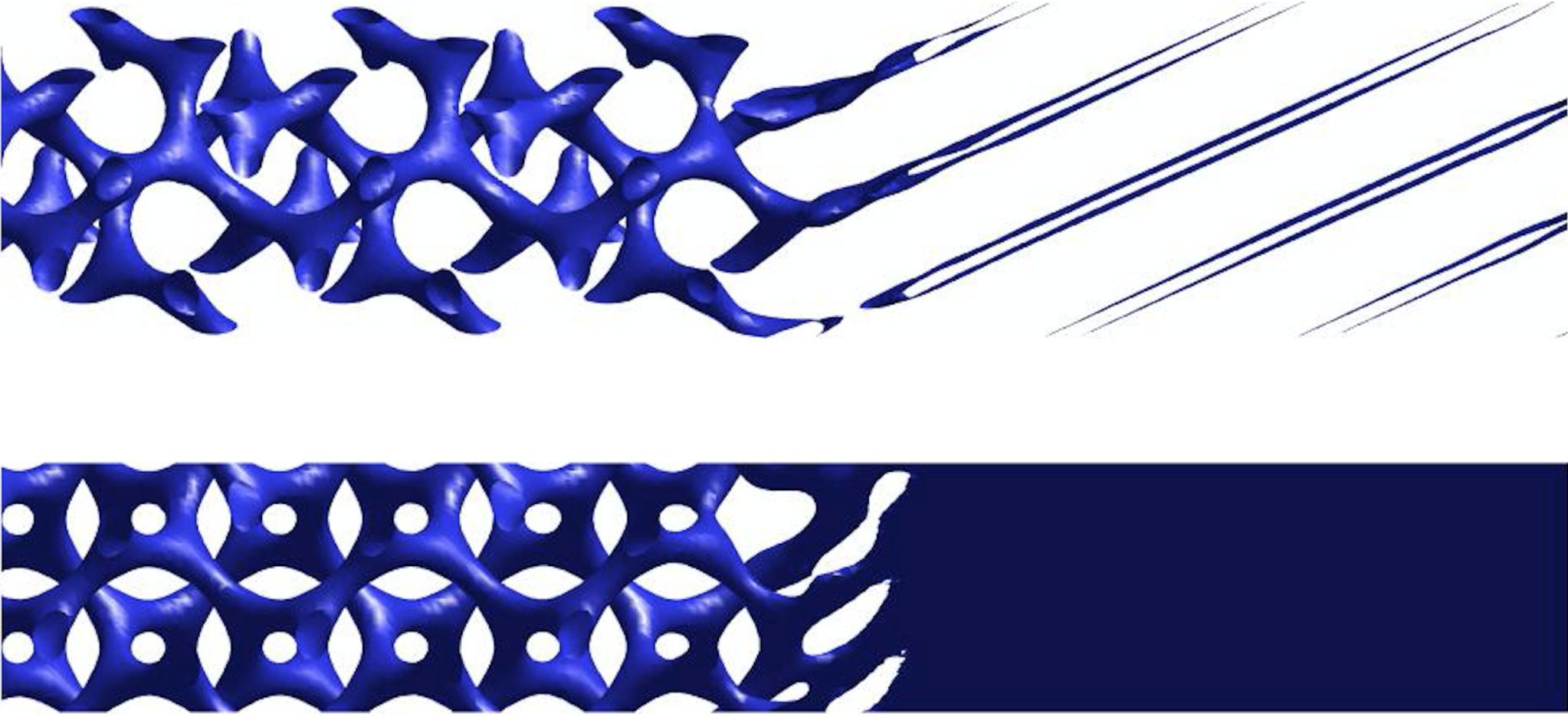}
  \includegraphics[width=0.48\textwidth,keepaspectratio]{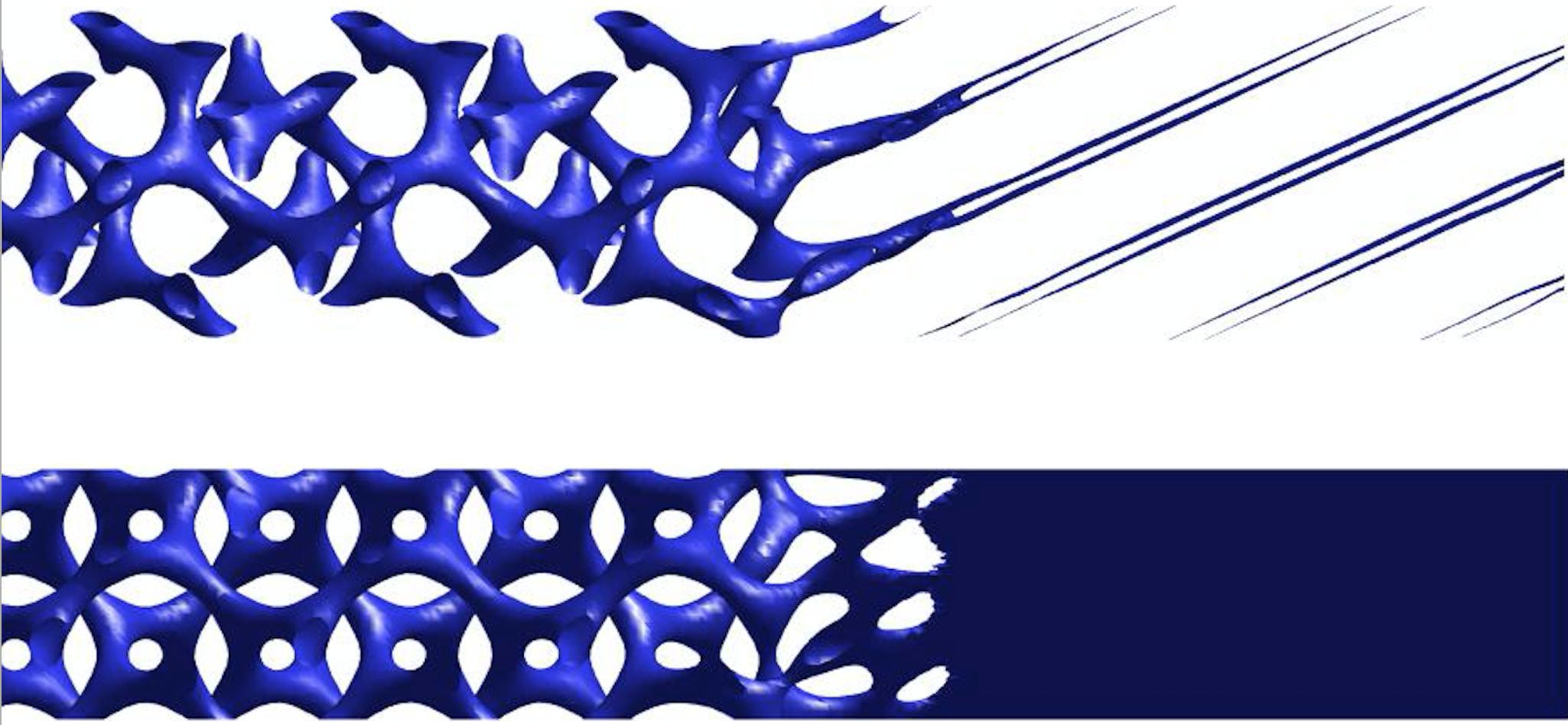}
  \caption{The L-G interface. Left: the epitaxially matching case; 
    Right: L is shifted a half period. Viewed along the layer of L (upper) 
    and the unit cell of G (lower). }
  \label{LGshift}
\end{figure}
\begin{figure}
  \centering
  \includegraphics[width=0.55\textwidth,keepaspectratio]{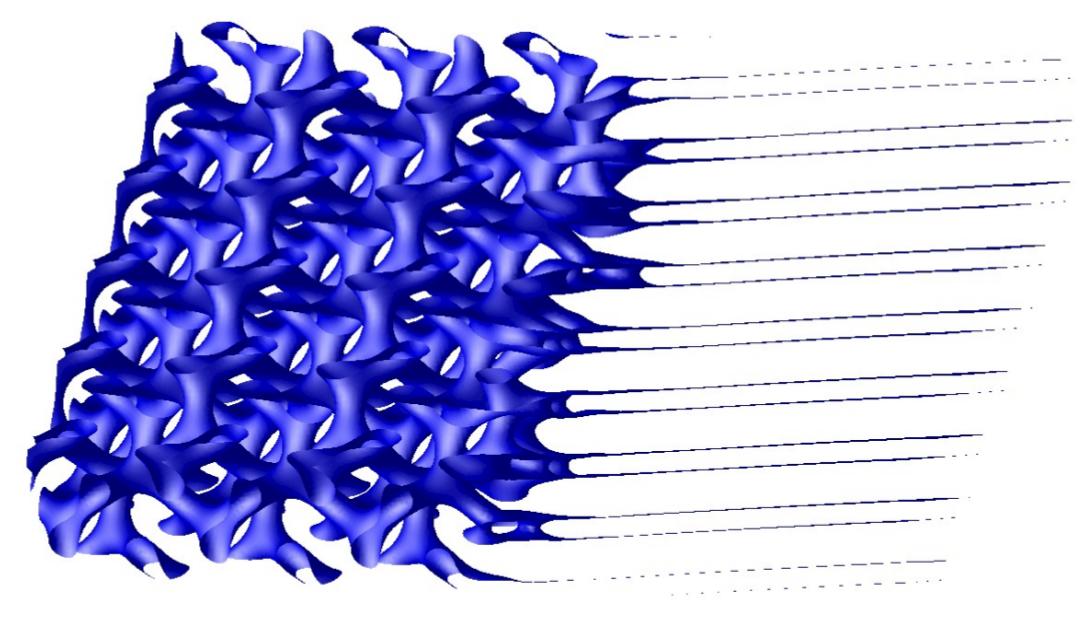}
  \includegraphics[width=0.4\textwidth,keepaspectratio]{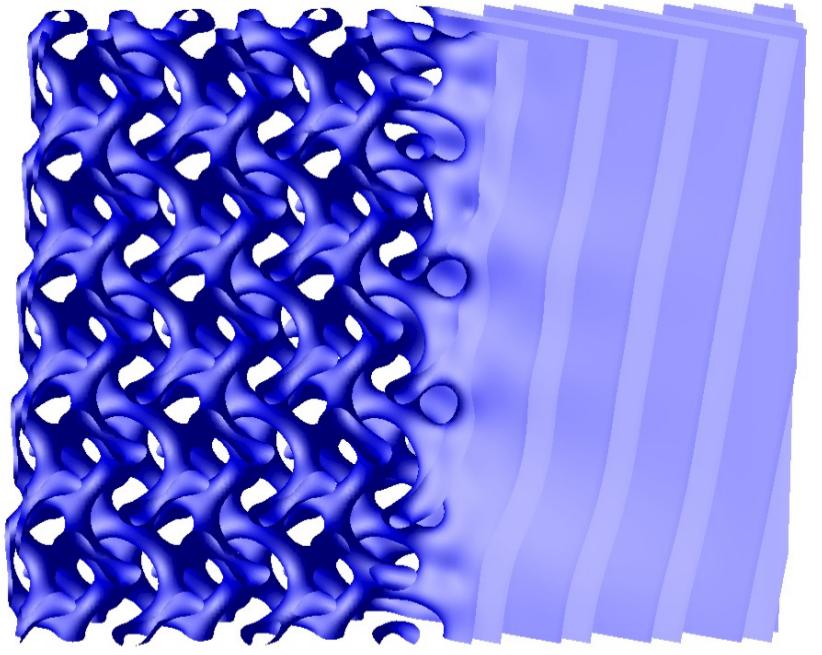}
  \caption{L-G interfaces with L rotated counter-clockwise 
    $\theta=\arcsin(3/5)$ (left), and $\theta+\pi/2$ (right). }
  \label{LGrotate}
\end{figure}
In the following, we will present some results of the non-matching cases, 
in which we can observe some interesting phenomena. 
The results described in this paragraph are calculated with 
$\xi^2=0.0389$, $\gamma=0.0601$, $\tau=-0.0121$. 
First, we examine the case where L is shifted a half period 
(Fig. \ref{LGshift} right). 
Comparing it with the epitaxially matching case (Fig. \ref{LGshift} left), 
we find the structure distinct from L and G in the middle. 
It resembles the metastable perforated layer structure, 
supporting the prevalent observation of perforated layer phase in the 
L$\leftrightarrow$G transitions (see the discussion in \cite{PRL_Ncl}). 
Next we look at the effects of relative rotation. 
In Fig. \ref{LGrotate} L is rotated $\theta=\arcsin(3/5)$ and 
$\theta+\pi/2$ counter-clockwise respectively, G unchanged. 
Local distortions help to keep their connection, leading to non-planar 
interfaces. 

\begin{figure}
  \centering
  \includegraphics[width=0.48\textwidth,keepaspectratio]{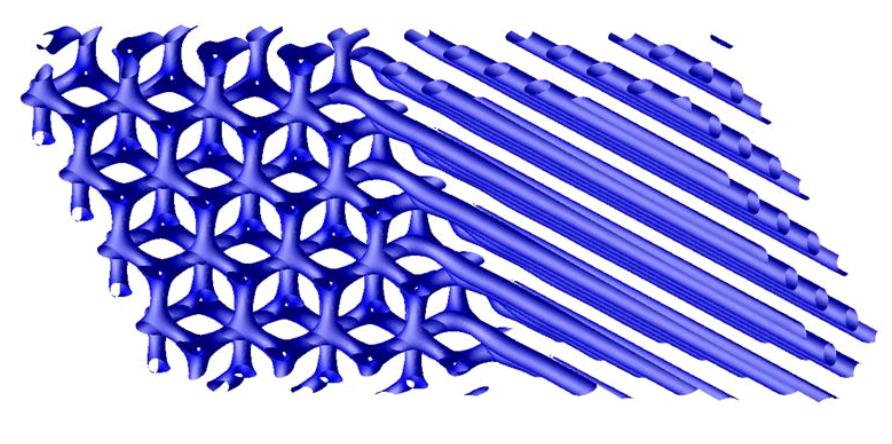}
  \includegraphics[width=0.45\textwidth,keepaspectratio]{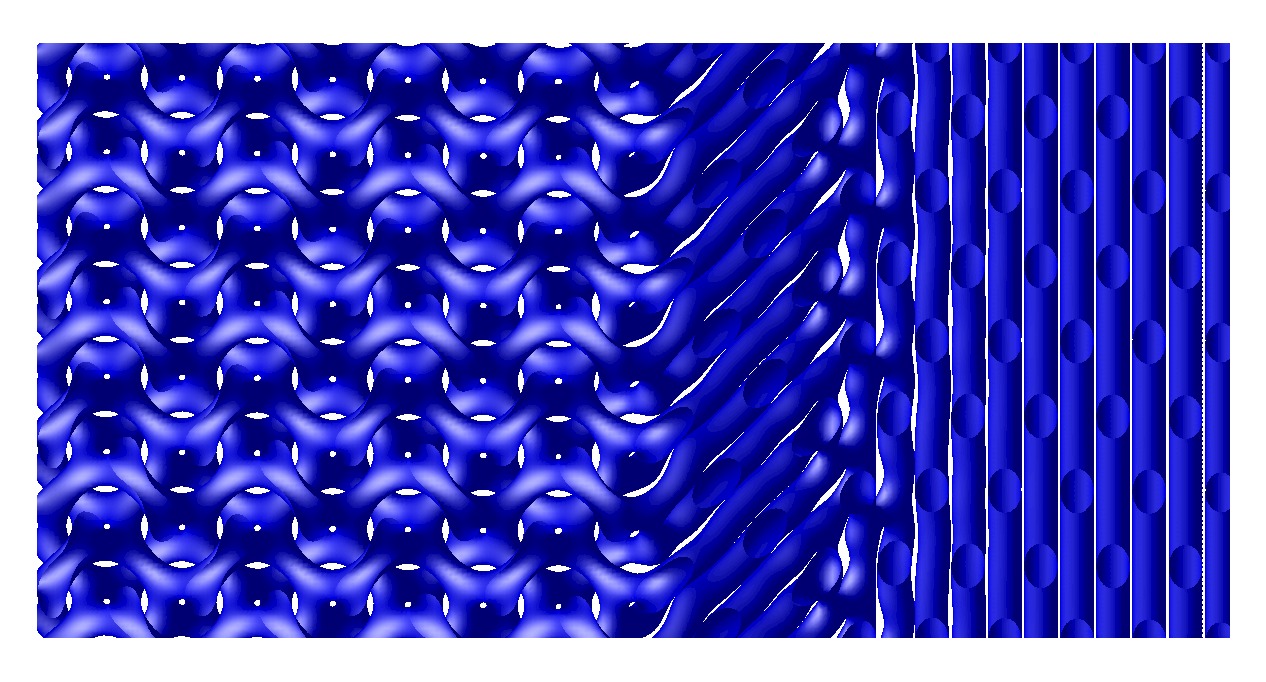}
  \caption{Left: C-G interface with newly found epitaxy $(1\bar{1}0)$. 
    Right: C is rotated $\pi/2$ from the new epitaxy and  
   cylinder along $\left<111\right>$ direction is found in between.}
  \label{CGnew}
\end{figure}

The next pair of examples are based on the newly-found epitaxially 
relationship between C and G \cite{JACS2009}, where the lattice of C is 
slightly deformed from the regular hexagon and lies in the plane $(1\bar{1}0)$. 
The parameters are chosen as $\xi^2=0.0375$, $\gamma=0.0757$, $\tau=-0.0102$. 
The interface is drawn in the left of Fig. \ref{CGnew}, 
which is planar with smooth connection. 
The right of Fig. \ref{CGnew} shows the interface 
where C is rotated $\pi/2$ in the $x$-$y$ plane. 
To connect two phases, C of the regular hexagonal type with classical epitaxy 
appears in between. 

\section{Summary\label{concl}}
A general framework is proposed for the computation of interfacial structures 
between modulated phases. 
The boundary conditions and the basis functions are carefully chosen to 
anchor the bulk phases with given position and orientation with compatibility. 
Because of the anistropy in the modulated structure, no extra 
constraint is necessary to stabilize the interface. 
In this way, the optimal interfacial structure is posed as a minimization 
problem that enables us to reach full relaxation and utilize fast optimization 
methods. We apply the framework to Landau-Brazovskii model. 
L-G and C-G interfaces with various relative positions and orientations 
are investigated, where some complex structures are obtained. 

So far we show the application of the framework to special cases where public 
period can be found in the $y$-$z$ plane. The choice of basis functions 
(\ref{basis}) actually allows us to investigate quasiperiodic phases. 
Thus in the future we aim to apply this framework to broader cases, 
especially for quasiperiodic phases. 
\vspace{20pt}

\textbf{Acknowldegment}
Pingwen Zhang is partly supported by National Natural Science Foundation of 
China (Grant No. 11421101, No. 11421110001 and No. 21274005).

\appendix
\section{Numerical details}
We describe some numerical details of discretization and optimization. 
For the discretization of density profile, 
finite difference scheme is adopted in the $x$-direction. In the $y$-$z$ plane, 
both finite difference scheme and Fourier expansion can be used. 

In the $x$-direction, Laplacian is approximated by
$$
\partial_x^2\phi(x_k)\approx\delta_x^2\phi(x_k)=
\frac{\phi_{k+1}-2\phi_k+\phi_{k-1}}{\Delta x^2}. 
$$
The same approximation is adopted when using finite difference scheme in 
the $y$-$z$ plane. 
When using Fourier expansion in the $y$-$z$ plane, we write $\phi$ as 
$$
\phi(x,y,z)=\sum_{\bm G} \phi_{\bm G}(x)\exp(i({\bm G}\cdot{\bm r}')). 
$$
where ${\bm G}=m{\bm b}_1'+n{\bm b}_2',\quad |m|,|n|\le N$, 
$\bm{b}_i'$ are reciprocal vectors with respect to the lattice in the $y$-$z$ 
plane, and ${\bm r}'=(y,z)$. Note that $\phi$ is real-valued, thus
it requires $\phi_{-{\bm G}}(x)=\phi^*_{\bm G}(x)$. 

The anchoring boundary conditions at $x=\pm L$ are implemented by adding two 
extra grids on each side and setting $\phi$ equal to bulk values. This can be 
equivalently viewed as approximating boundary conditions with finite difference 
scheme, 
$$
\phi(-L)=\phi_{\alpha}(-L), \qquad 
\partial_x\phi(-L)\approx\frac{\phi_{0}-\phi_{-1}}{\Delta x}
=\frac{\phi_{\alpha}(x_{0})-\phi_{\alpha}(x_{-1})}{\Delta x}
\approx\partial_x\phi_{\alpha}(-L).
$$

The gradient vector can be calculated by 
$$
\nabla \tilde{F}(\phi(x))=\xi^2[(\delta_x^2+\delta_y^2+\delta_z^2+1)^2+\tau]\phi(x) 
-\frac{\gamma}{2}\phi^2(x)+\frac{1}{6}\phi^3(x)
$$
for finite difference method, and by
\begin{equation}
\nabla \tilde{F}(\phi_{\bm G}(x))=\xi^2[(\delta_x^2-{\bm G}^2+1)^2+\tau]\phi_{\bm G}
-\frac{\gamma}{2}\sum_{{\bm G}_1+{\bm G}_2={\bm G}}\phi_{{\bm G}_1}
\phi_{{\bm G}_2}+\frac{1}{6}\sum_{{\bm G}_1+{\bm G}_2+{\bm G}_3={\bm G}}
\phi_{{\bm G}_1}\phi_{{\bm G}_2}\phi_{{\bm G}_3} \label{EL}
\end{equation}
for Fourier expansion. The convolution sum can be calculated by FFT. 

The conservation of $\phi$ is attained by a projection on the gradient vector: 
for finite difference scheme, we use 
$$
\nabla F(\phi(x))=\nabla\tilde{F}(\phi(x))-c; 
$$
and for Fourier expansion, we use 
$$
\nabla F(\phi_{\bm G}(x))=\nabla \tilde{F}(\phi_{\bm G}(x))-c\delta({\bm G}=0). 
$$
In the above, $c$ can be determined by the following observation: 
if we set $\phi=\phi_{\alpha}$ for $x<0$ and $\phi=\phi_{\beta}$ for $x>0$, 
the constraint is satisfied. So we can just require 
$$
\int_S\int_{-L}^L\phi \md x\md y\md z=
\int_S\md y\md z(\int_{-L}^0\phi_{\alpha}\md x+\int_0^L\phi_{\beta}\md x)
\triangleq c_0. 
$$

For finite difference scheme, we use a gradient method
$$
\phi^{n+1}(x)- \phi^n(x)=-a_n\nabla F(\phi^n(x)). 
$$
The coefficient $a_n$ is altered adaptively with Barzilai-Borwein method
\cite{BB0,BB}. 
For Fourier expansion we use a semi-implicit scheme to solve 
the Euler-Langrange equation (\ref{EL}) with 
\begin{eqnarray*}
\frac{\phi^{n+1}_{\bm G}(x)-\phi^{n}_{\bm G}(x)}{\Delta t}&=&
-\xi^2[(\delta_x^2-{\bm G}^2+1)^2+\tau]\phi^{n+1}_{\bm G}\\
&&+\frac{\gamma}{2}\sum_{{\bm G}_1+{\bm G}_2={\bm G}}\phi^n_{{\bm G}_1}
\phi^n_{{\bm G}_2}-\frac{1}{6}\sum_{{\bm G}_1+{\bm G}_2+{\bm G}_3={\bm G}}
\phi^n_{{\bm G}_1}\phi^n_{{\bm G}_2}\phi^n_{{\bm G}_3}\\
&&-c\delta({\bm G}=0).
\end{eqnarray*}

\bibliographystyle{unsrt}
\bibliography{alg_bib} 

\end{document}